# Thermodynamics and Kinetics of Ti$_2$N Formation by N Atom Intercalation in Ti

J. Varalakshmi[1,2] and Satyesh Kumar Yadav[1,2,*]

[1] *Department of Metallurgical and Materials Engineering, Indian Institute of Technology (IIT) Madras, Chennai 600036, India.*

[2] *Center for Atomistic Modelling and Materials Design, Indian Institute of Technology (IIT) Madras, Chennai 600036, India.*

*Corresponding author: satyesh@iitm.ac.in

**Abstract**

Although Ti$_2$N and Ti belong to different crystal systems and Bravais lattices, we show that N intercalation in Hexagonal Close-Packed (HCP) Ti can lead to the formation of Ti$_2$N by diffusionless displacement of Ti atoms, which is a nucleation-free growth process. This is due to the structural similarity of Ti atoms in Ti and Ti$_2$N. For N intercalation in Ti, the first N$_2$ molecule should adsorb on the surface of Ti {0001} and dissociate into N atoms, and then enough of the adsorbed N atoms should diffuse into Ti to form Ti$_2$N. We calculated the enthalpy of formation for each step during N intercalation using density functional theory (DFT). We found that N intercalation was thermodynamically favored at each step of nitridation till the formation of Ti$_2$N. We identified the transition network for each diffusion path and calculated the diffusion coefficient of N from surface to sub-surface to bulk.

## 1. Introduction

Range of Ti and N compounds like TiN, Ti$_6$N$_5$, Ti$_4$N$_3$, Ti$_3$N$_2$, Ti$_2$N [1–4] and solid solutions of nitrogen in titanium (~ 17-23 at% N depending on temperature) [3–5] are stable and have been synthesized. Ti$_2$N forms from Ti in the presence of N, especially when the concentration of N is insufficient to form TiN. Ti$_2$N crystal growth was noticed during reactive magnetron sputtering of Ti in the presence of low N flow [6]. It also appears when Ti is subjected to nitriding at temperatures between 800-1200 °C [7,8]. Titanium ion nitride in a nitrogen plasma at 800 $^0$C, forms Ti$_2$N [9]. Additionally, Ti$_2$N forms during the deposition of Ti via cathode arc evaporation in the presence of N$_2$ gas [10]. Ti$_2$N also forms at the interface of Ti and TiN during the growth of Ti/TiN multilayer [11].

The formation of Ti$_2$N from Ti under low concertation of N is expected due to its thermodynamic stability. However, the formation of Ti$_2$N by direct diffusion of N in a Ti is neither expected nor explored. Ti$_2$N belongs to P4$_2$/mnm space group and tetragonal crystal system [12], while Ti belongs to P6$_3$/mmc space group and hexagonal crystal system [13]. So, one can expect a rearrangement of the Ti atoms in the Ti hexagonal crystal under low concertation of N to form Ti$_2$N. These atom rearrangements should require nucleation of Ti$_2$N in the Ti, followed by growth of Ti$_2$N. This study explored the possibility of N intercalation in the Ti to form Ti$_2$N, which would be a nucleation-free growth process.

## 1.1 Structural comparison of Ti and Ti$_2$N structures

Fig. *1* shows the atomic arrangement of Ti atoms in both Hexagonal Close-Packed (HCP) Ti (Fig. 1(a)) and Ti$_2$N tetragonal (Fig. 1(b)) structures. The stacking sequence of Ti atoms in Ti$_2$N along the [010] direction is similar to the HCP Ti, ABAB stacking sequence along the [0001] direction. When comparing the atomic positions of Ti atoms in the two structures, Ti1 atoms occupied corner sites in both structures (Fig. 1(c) and (d)). Ti2 atom in Ti$_2$N was slightly shifted downwards from the face-centered position compared to the HCP Ti (Fig. 1(c) and (d)). Similarly, Ti3 and Ti4 atoms in Ti$_2$N were slightly displaced from their positions compared to the HCP Ti. Ti-Ti bond length in Ti$_2$N along the [010] direction was 6% larger than Ti-Ti bond length in HCP Ti along the [0001] direction (Fig. 1(c) and (d)). Ti-Ti bond length in Ti$_2$N along [100] direction was 3% shorter, and the same was 3% larger along [001] direction compared to Ti-Ti bond lengths in HCP Ti along [1$\bar{2}$10] and [10$\bar{1}$0] directions, respectively (Fig. 1(e)-(f)). These observations suggest that nitrogen can intercalate into the Ti lattice to form Ti$_2$N by a small movement of Ti atoms and a small change in HCP Ti volume. These details provide valuable insights into the structural changes occurring when N moves in Ti to form Ti$_2$N.

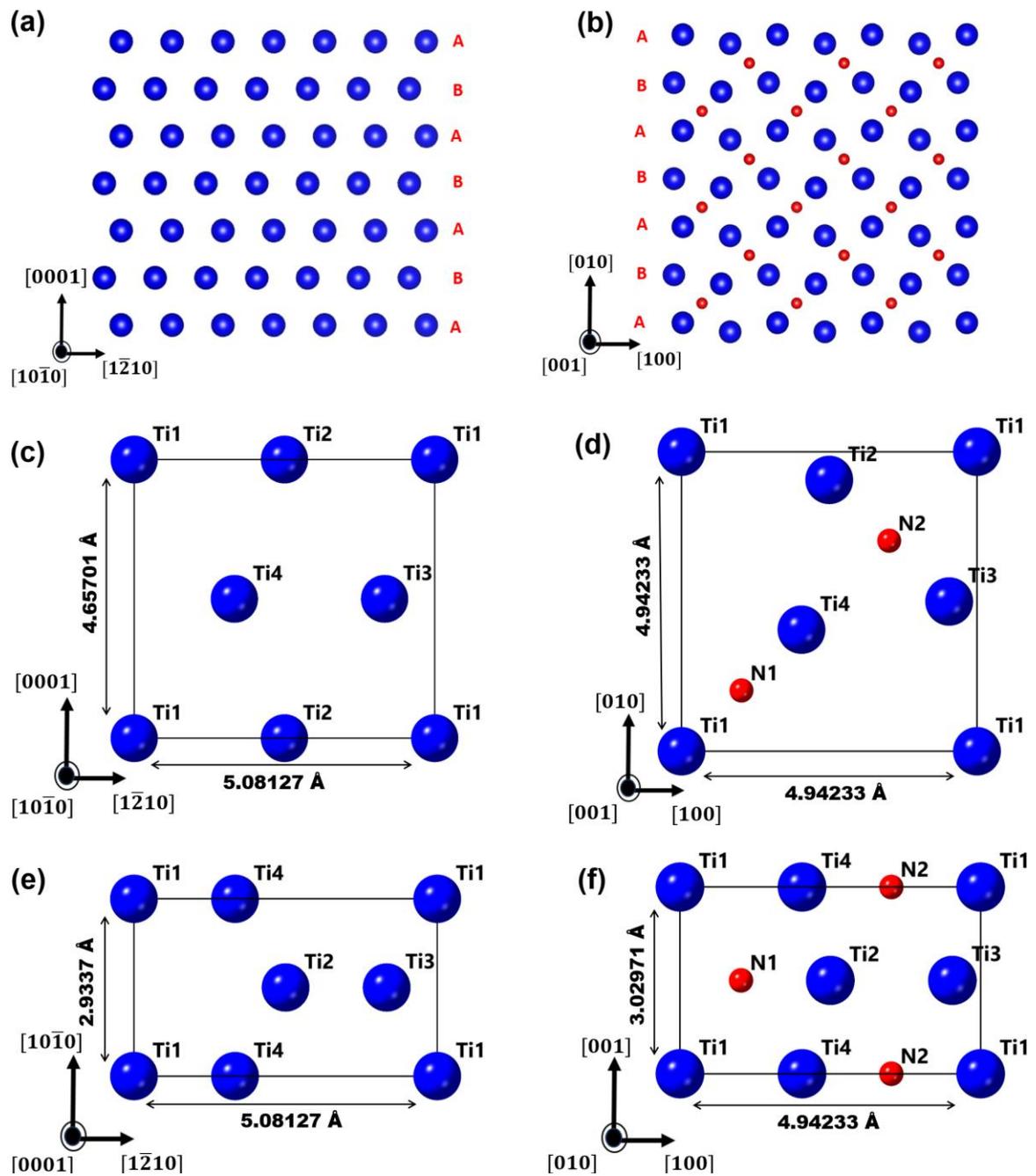

Fig. 1. Atomic structure comparison between Ti and Ti$_2$N (a) and (b) show the stacking sequence of the HCP Ti and Ti$_2$N tetragonal structures along the [0001] and [010] directions, respectively. Atomic arrangements for Ti1, Ti2, Ti3, and Ti4 atom positions in the Ti and Ti$_2$N structures are shown in (c)- (f). Here, (c) & (e) refer to the Ti system, and (d) & (f) refer to the Ti$_2$N system. The red-coloured sphere represents the N atom, and the blue-coloured sphere represents the Ti atom.

To form a Ti$_2$N from the Ti by intercalating N atoms, N$_2$ molecules first need to be adsorbed on the Ti surface, then split into N atoms, followed by diffusion of the N atoms into the sub-surface and subsequently into the bulk region of the Ti system. Each step of nitridation of Ti should be thermodynamically favored. Also, subsequent steps should be more favored than the previous step. The rate of transformation from any step to the next step should be fast enough to observe the transformation in real time. In order to access the thermodynamic stability of various steps and rate of nitridation till the formation of Ti$_2$N, we performed atomistic modelling using density functional theory (DFT) of (1) Adsorption of the N$_2$ molecule on the Ti {0001} surface, (2) N atom adsorption on the Ti {0001} surface by splitting of the adsorbed N$_2$ molecule, (3) N intercalation at the sub-surface Ti{0001}, (4) N intercalation in the bulk Ti, (5) formation of N solid solution in Ti, and finally (6) formation of Ti$_2$N.

## 2. Methodology

The first-principles atomistic modelling calculations were conducted using DFT [14,15] as implemented in the Vienna Ab initio Simulation Package (VASP) code [16–19]. The DFT calculations utilized the Perdew, Burke, and Ernzerhof (PBE) [20] generalized gradient approximation (GGA) exchange-correlation functional and the projector-augmented wave (PAW) method [21,22]. A plane wave kinetic energy cutoff of 520 eV was used for the plane wave expansion of the wave functions to achieve highly accurate forces. The Brillouin zone sampling was performed using Monkhorst-Pack meshes [23] with a resolution of less than 2π x 0.03 Å$^{-1}$. The structures were relaxed using a conjugate gradient method, with total energies converged to within 1meV per atom. Similar energy convergence was achieved in tests using 4 x 4 x 3 supercells containing (1) 96 Ti atoms, (2) 96 Ti atoms with 1 N atom (Ti0.01N), and (3) 96 Ti atoms with 24 N atoms (Ti-0.20N), using K points of 2 x 2 x 2. For the 2 x 2 x 3 supercell with 48 Ti atoms and 24 N atoms, K points of 5 x 5 x 5 were used. Slabs containing (1) N$_2$ molecule in 63 Ti atoms, (2) Single N in 64Ti atoms with K points of 5 x 5 x 1, 10 Å vacuum were used, respectively, and (3) 126 Ti atoms and 24 N atoms of Ti/Ti$_2$N with K-points 1 x 1 x 1, 10 Å vacuum were used for determine adsorption energy, relaxation calculations. We used the climbing-image nudged elastic band (C-NEB) [24,25] method with three intermediate images and maintained constant cell shape to identify the transition pathways and energy barriers between the initial and final position of hopping N. The forced component is negated along the path while the perpendicular components remain unchanged.

The images relax to an extremum where the forces are less than 5meV/ Å, and the restoring force verifies that this extremum is a first-order saddle point.

## 3. Results and discussion

The step-by-step formation of the $Ti_2N$ structure through nitrogen intercalation in the Ti is schematically shown in Fig. 2. First, $N_2$ molecules adsorbed onto Ti's {0001} surface (at layer 6), specifically at the top of the FCC sites (Top view-I image in Fig. 2), since it was reported to be the most stable site [26]. These $N_2$ molecules subsequently dissociated into individual N atoms, occupying the FCC sites on the topmost layer of the Ti {0001} surface (Top view-II image in Fig. 2). Nitrogen atoms then migrated into the sub-surface region and occupied the octahedral sites (Top view-III image in Fig. 2). More nitrogen migration into Ti leads to the formation of various N-solid solutions. Once enough N atoms have diffused into Ti, a $Ti_2N$ crystal can form, as shown in the subfigure (at the bottom of the supercell) in Fig. 2. We discussed the thermodynamics and kinetics favouring this process in the following subsections.

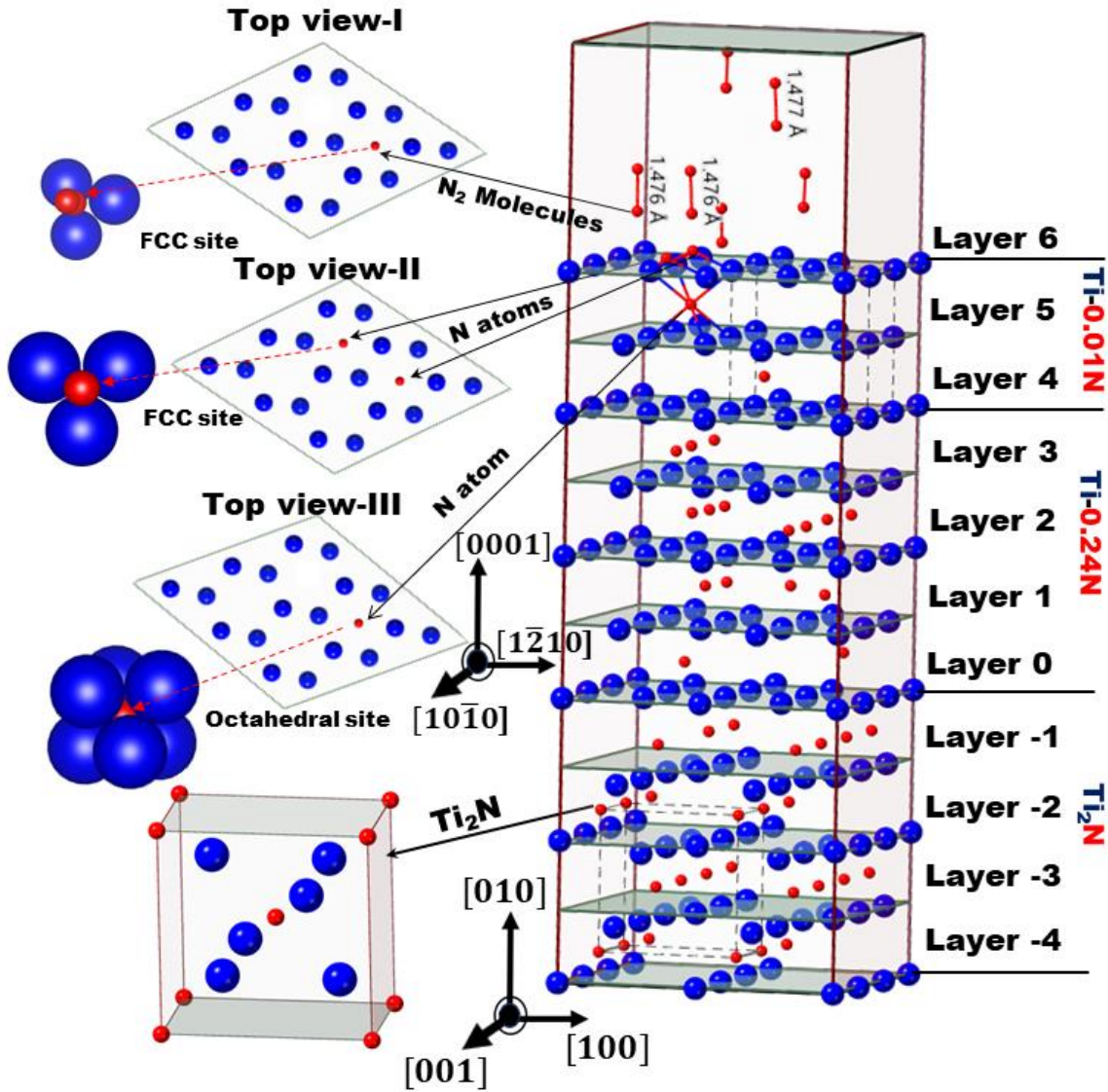

Fig. 2. Schematic of the formation of Ti$_2$N from Ti. Layer 0 represents the interface between Ti and Ti$_2$N. Layers -1 to -4, 1 to 3, and 4 to 6 represent Ti$_2$N, Ti-0.20N, and Ti-0.01N structures, respectively. Top view-I: N$_2$ molecules adsorption. Top view-II: N atoms at FCC site. Top view-III: N atom at octahedral site; Subfigure at the bottom: Ti$_2$N.

### 3.1 Thermodynamics of N atom intercalation in Ti

In order to access the thermodynamic stability of each step (described previously) and their relative stability, we calculated the enthalpy of formation per N atom in each step with Ti bulk and N$_2$ molecules as references. The equations for calculating the formation enthalpy of various steps were described as follows. N$_2$ molecules are placed on the top of FCC sites in the first step, as shown in Fig. 3(a). Adsorption energy per N₂ molecule on Ti {0001} surface, $\Delta H_f(N_2^{ads} on Ti^{surf})$, is given by equation (1).

$$\Delta H_f(N_2^{ads}\ on\ Ti^{surf}) = \frac{E_{slab}(N_2^{ads} on\ Ti^{surf}) - E_{slab}(Ti^{surf}) - E(N_2)}{2} \quad (1)$$

where, $E_{slab}(N_2^{ads} on\ Ti^{surf})$ is the DFT total energy of the N₂ molecule adsorbed on a 63 Ti atom slab, as shown in Fig. 3(a). The slab (Fig. 3(a)) contains 7 Ti closed-pack plane layers with 9 Ti atoms in each layer. $E_{slab}(Ti^{surf})$ is the DFT total energy of 63 Ti atoms slab and $E(N_2)$ is the DFT total energy of the N₂ molecule.

In the next step, N atoms were placed at the FCC sites on both the top and bottom surfaces of the Ti {0001} surface, as shown in Fig. 3(b). Adsorption energy per N atom on the FCC site of Ti {0001} surface, $\Delta H_f(N^{ads} on Ti_{FCC}^{surf})$, is given in equation (2).

$$\Delta H_f(N^{ads}\ on\ Ti_{FCC}^{surf}) = E_{slab}(2N^{ads}\ on\ Ti_{FCC}^{surf}) - E_{slab}(Ti) - 2\ E(N_2) \quad (2)$$

where, $E_{slab}(2N^{ads} on Ti_{FCC}^{surf})$ is the DFT total energy of 2N atoms adsorbed on the 63 Ti atom slab (Fig. 3(b)).

In the following step, the N atom was placed at the sub-surface octahedral site below the Ti {0001} surface, as shown in Fig. 33(c). Absorption energy per N atom at the sub-surface octahedral site, $\Delta H_f(N^{abs}\ on\ Ti_{O_{sub}}^{surf})$, is given in equation (3).

$$\Delta H_f(N^{abs}\ on\ Ti_{O_{sub}}^{surf}) = E_{slab}(N^{abs}\ on\ Ti_{O_{sub}}^{surf}) - E_{slab}(Ti) - \frac{1}{2}\ E(N_2) \quad (3)$$

where, $E_{slab}(N^{abs} on\ Ti_{O_{sub}}^{surf})$ is the DFT total energy of the N atom absorbed on the sub-surface octa site in the 63 Ti atom slab.

In the subsequent step, the N atom was placed at the octahedral site, as shown in Fig. 3(d). The formation energy per N interstitial in the HCP Ti, $\Delta H_f(Ti - 0.01N)$, is given in equation (4).

$$\Delta H_f(Ti - 0.01N) = E_{bulk}(Ti - 0.01N) - E_{bulk}(Ti) - \frac{1}{2}\ E(N_2) \quad (4)$$

where, $E_{bulk}(Ti - 0.01N)$ is the DFT total energy of a 96 Ti atoms slab with 1 N interstitial at the octahedral site. $E_{bulk}(Ti)$ is the DFT total energy of the 96 Ti atoms slab. Here, we started with a primitive HCP Ti cell, which contained two atoms and replicated as 4 x 4 x 3 times along the X, Y, and Z directions to form a supercell with 96 Ti atoms.

The formation energy per N atom in N-solid solution of bulk Ti, $\Delta H_f(Ti - 0.20N)$, in the following step, is given in equation (5).

$$\Delta H_f(Ti - 0.20N) = \frac{E_{bulk}(Ti - 0.20N) - E_{bulk}(Ti) - 12(N_2)}{24} \qquad (5)$$

where, $E_{bulk}(Ti - 0.20N)$ is the DFT total energy of the 96 Ti atoms with 24 N atoms at the octahedral site. To create a Ti-0.20N supercell, we used the 96 Ti atoms supercell and introduced 24 N atoms into the octahedral sites, ensuring the maximum average distance between the N atoms. One N atom was initially placed within the Ti supercell in a randomly selected octahedral void (first void). The next octahedral void (second void) chosen was the one furthest from the first void, and the third void was selected to be the farthest from both the first and second voids. This process is continued to fill subsequent voids, as shown in Fig. 3(e).

Finally, Fig. 3(f) presents the modelled Ti$_2$N structure with a space group of P4$_2$/mnm. The formation energy of Ti$_2$N, $\Delta H_f(Ti_2N)$, is given in equation (6).

$$\Delta H_f(Ti_2N) = E_{bulk}(Ti_2N) - E_{bulk}(Ti) - \frac{1}{2}E(N_2) \qquad (6)$$

where, $E_{bulk}(Ti_2N)$ is DFT total energy per formula unit of Ti$_2$N, $E_{bulk}(Ti)$ is DFT total energy per bulk Ti atom.

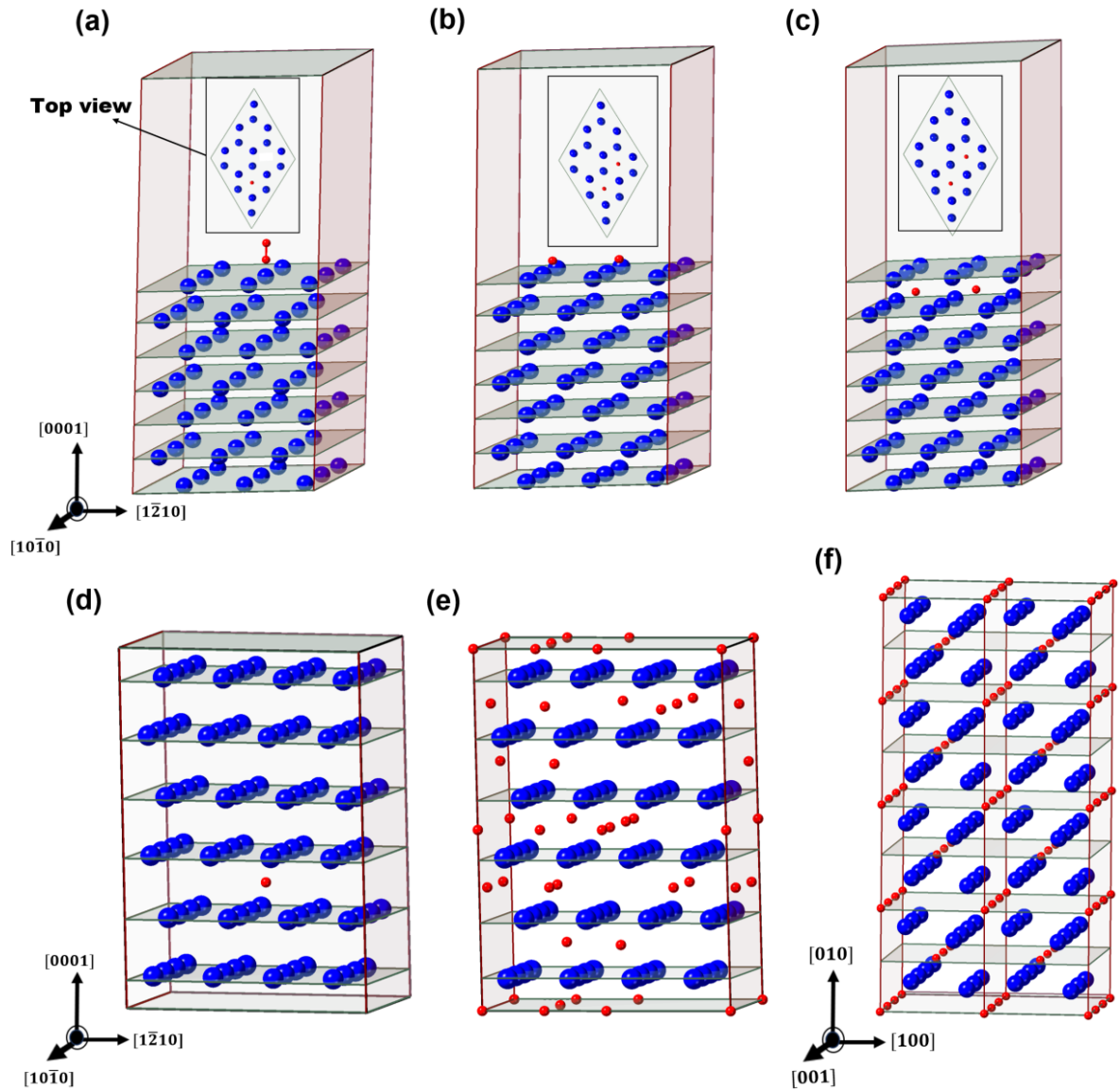

Fig. 3. Atomic structures of (a) $N_2$ adsorption on {0001} surface (b) N atoms adsorption on {0001} surface c) N atom in sub-surface in Ti {0001} and N concentration in bulk Ti {0001}: (d) Ti-0.01N e) Ti-0.20N and f) $Ti_2N$. The inscribed figures in (a), (b), and (c) show the top view.

The calculated enthalpies of formation per N atom using equations (1) to (6) for various intermediate structures in the nitridation of Ti are listed in Table 1. It is evident from Table 1 that each intermediate step has a negative enthalpy of formation. The enthalpy of the formation of the structure in the next step was more negative than that of the structure in the current step. Therefore, $N_2$ molecules could adsorb on the surface. The dissociation of $N_2$ molecules into N atoms (Fig. 3(b)) further reduced the enthalpy of formation, favouring the process.

Similarly, the N atom could diffuse into Ti bulk to form a solid solution as the enthalpy of formation per N of the solid solution was more negative than the adsorption energy of the N atom on the Ti {0001} surface. Finally, Ti$_2$N could form when sufficient N atoms have been absorbed into the bulk Ti, which had the most negative enthalpy of formation. Thus, it was observed that thermodynamics favoured the proposed process for Ti$_2$N formation.

Table 1. Comparison of adsorption energy of N$_2$ molecule and N atom on Ti {0001} surface and formation energy of solid solution of N in Ti and Ti$_2$N.

| N atom Adsorption on various intermediate structures | Energy (eV/N atom) |
|---|---|
| N$_2$ on {0001} Ti | -0.31 |
| N on {0001} Ti | -2.87 |
| N absorption at sub-surface | -3.30 |
| Ti-0.01N | -3.71 |
| Ti-0.20N | -3.83 |
| Ti$_2$N | -3.99 |

## 3.2 Kinetics of N atoms intercalation in Ti

To understand the rate-limiting step in the intercalation of N atom in Ti to form Ti$_2$N, we calculated the diffusivity of (1) N diffusion from Ti {0001} surface to sub-surface (2) N diffusion from Ti {0001} sub-surface to Ti bulk (3) N diffusion in HCP Ti with 1% N interstitial (Ti-0.01N) (4) N diffusion in 20% N solid solution in Ti (Ti-0.20N) (5) last N atom diffusion to form Ti$_2$N. The transition rate from the initial equilibrium position, *i,* to the final equilibrium position, *j,* at temperature *T* is an Arrhenius equation, given by the following equation (7) [27,28].

$$\lambda_{ij} = v_{ij}\exp(\frac{-E_{ij}}{k_B T}) \qquad (7)$$

where, $\lambda_{ij}$ is the jump rate from *i* to *j* and $E_{ij}$ determines the barrier energy of the system. This barrier energy represents the difference between the initial equilibrium lattice position of the diffusing atom (initial state) and the saddle point along the diffusion path (transition state). $K_B$ and $T$ denote the Boltzmann constant ($8.617330 \times 10^{-5}$ eV/K) and the absolute temperature. $v_{ij}$ is the attempt frequency prefactor for the transition. It is a ratio of the product of all *3N* vibrational modes at the initial state (non-imaginary modes at minima) to the product of all *3N-1* vibrational modes at the transition state (saddle point). The attempt frequency prefactor for each transition is estimated in the Vineyard equation (8) [28,29].

$$v_{ij} = \frac{\prod_i^{3N} v_i^{minima}}{\prod_j^{3N-1} v_j^{saddle}}) \qquad (8)$$

where, $v_i^{minima}$ is attempt frequency at minima and $v_j^{saddle}$ is attempt frequency at the saddle point. We identified possible diffusion paths for each step and calculated the corresponding barrier and attempt frequency. These details were discussed in the following subsections.

### *3.2.1 N atom diffusion from {0001} Ti surface to sub-surface*

As mentioned previously, the FCC site on the Ti {0001} surface and the octahedral site (just beneath the surface) in bulk Ti were the most stable sites for the N atom [30]. The diffusion path connecting the initial state (IS) (FCC site) and final state (FS) (sub-surface octahedral site) through a transition state (TS) is indicated in

Fig. 4. Barrier energy and attempt frequency prefactor for the same path are calculated and listed in Table 3. The diffusivity coefficient $D_{FCC \to O_{sub}}^{surface}$ for N atom diffusion from the surface FCC site to the octahedral site is calculated using the diffusion equation (9).

$$D_{FCC \to O_{sub}}^{surface} = \ell_{FCC \to O_{sub}}^2 (\lambda_{FCC \to O_{sub}}) \qquad (9)$$

where, $\ell_{FCC \to O_{sub}}$ is the diffusion length between the FCC site and sub-surface (FCC-$O_{sub}$). $\lambda_{FCC \to O_{sub}}$ is jump rate from the surface FCC site to the sub-surface octahedral (FCC-$O_{sub}$). The diffusivity coefficient at room temperature is listed in Table 3.

### *3.2.2 N diffusion from {0001} Ti sub-surface to bulk Ti*

For N diffusion from the sub-surface to a bulk-like region in Ti along $[000\bar{1}]$ direction, we identified two possible pathways: (1) sub-surface octahedral site to bulk octahedral site ($O_{sub}$-$O_{bulk}$), and (2) sub-surface octahedral site to sub-surface hexahedral site ($O_{sub}$-$H_{sub}$), from then to the bulk octahedral site ($H_{sub}$-$O_{bulk}$), as shown in

Fig. 4(b). Identified TSs for these transition paths by connecting ISs and FSs are indicated in

Fig. 4(b). As the N atom moves along $[000\bar{1}]$ direction, as in path-1, the N atom experiences three nearest Ti atoms (can be referred to as a triangular face), as shown in inscribed Fig. 4(b). The triangular face of three Ti atoms demands more energy to displace Ti atoms [31] in close-packed {0001} planes, leading to a higher barrier energy of 3.50 eV compared to the $O_{sub}$-$H_{sub}$ (path-2) barrier energy of 1.73 eV (Table 3). The barrier for $H_{sub}$-$O_{bulk}$ (0.48 eV (Table 3)) was substantially smaller than $O_{sub}$-$H_{sub}$. Hence, we consider only the O-H jump while ignoring the H-O jump to calculate the diffusion coefficient. This assumption is consistent with earlier reports. [28,32,33]. Therefore, the diffusivity of the N atom from $O_{sub}$ to $H_{sub}$ to $O_{bulk}$ could be dictated by the transition from $O_{sub}$ to $H_{sub}$. Considering both pathways, the effective diffusion coefficient for N atom diffusion from the Ti {0001} sub-surface octahedral site to the bulk octahedral site is given by equation (10).

$$D^{subsurface}_{O_{sub} \to O_{bulk}} = \ell^2_{O_{sub} \to O_{bulk}}(\lambda_{O_{sub} \to O_{bulk}} + \frac{3}{2}\lambda_{O_{sub} \to H_{sub}}) \qquad (10)$$

where, $D^{subsurface}_{O_{sub} \to O_{bulk}}$ is the effective diffusion coefficient of the $O_{sub}$-$O_{bulk}$, $\ell_{O_{sub} \to O_{bulk}}$ is the diffusion length of $O_{sub}$-$O_{bulk}$. $\lambda_{O_{sub} \to O_{bulk}}$, and $\lambda_{O_{sub} \to H_{sub}}$ are the jump rates from the $O_{sub}$-$O_{bulk}$ and $O_{sub}$-$H_{sub}$, respectively. As the barrier for the $O_{sub}$-$H_{sub}$ path is significantly lower than the $O_{sub}$-$O_{bulk}$ path, diffusion of N atom could be dominated by the $O_{sub}$ - $H_{sub}$ - $O_{bulk}$ path.

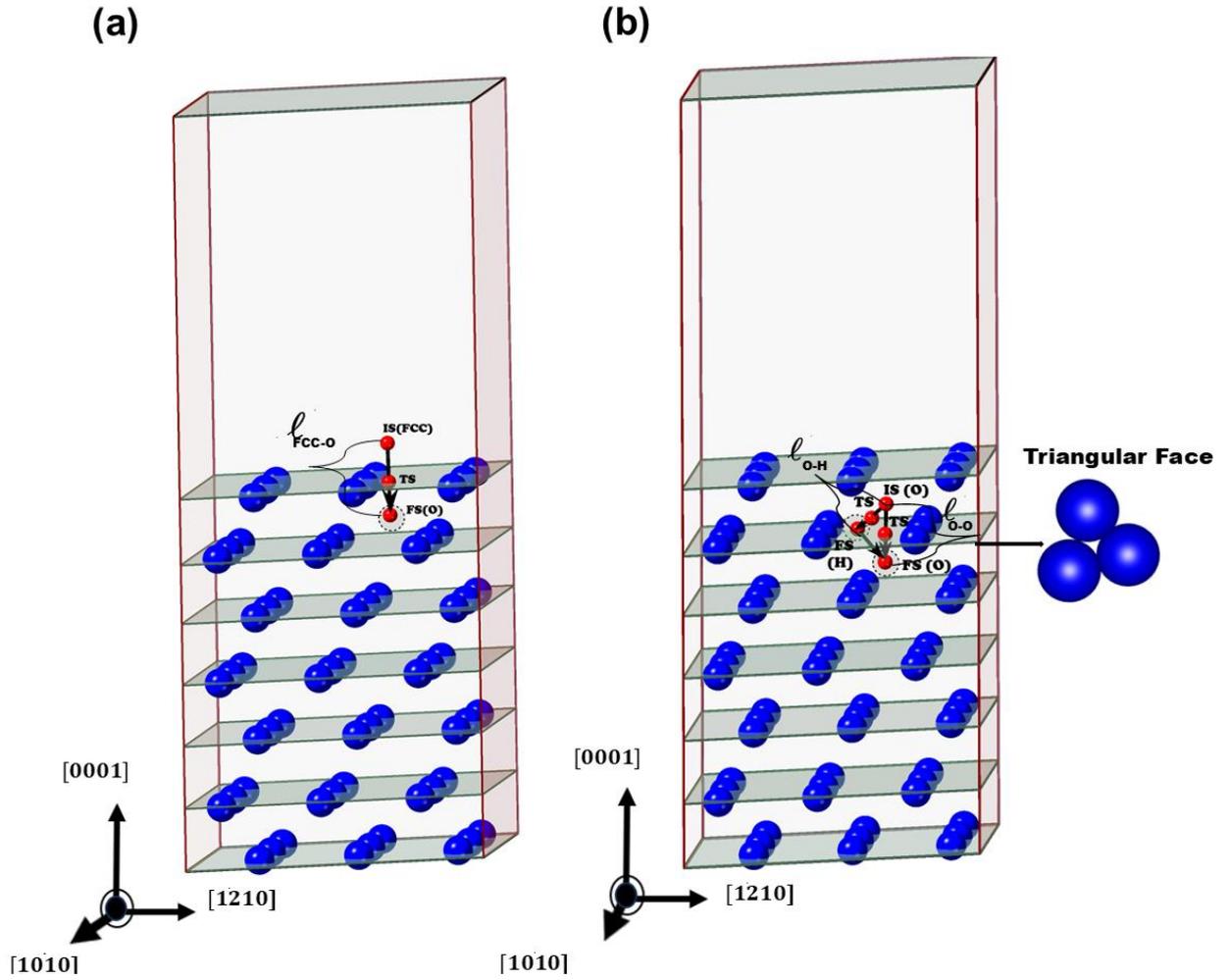

Fig. 4. N diffuses on Ti {0001} surface: (a) from FCC site to sub-surface octahedral site (FCC-$O_{sub}$), (b) from octahedral site to bulk octahedral site ($O_{sub}$-$O_{bulk}$) and then sub-surface octahedral site to sub-surface hexahedral site ($O_{sub}$ - $H_{sub}$). IS is an initial state, FS is a final state, and $\ell$ is diffusion length.

### 3.2.3 N diffusion in Ti-0.01N

N diffusion further into the bulk region of Ti along [000$\bar{1}$], the direction was modelled as 1% N interstitial diffusion in bulk HCP Ti. We found that N was stable at the hexahedral site. Thus, two diffusion paths were identified: (1) direct octahedral to octahedral (O-O) path and (2) octahedral to hexahedral to octahedral path (O-H-O). The diffusion paths and TSs are shown in Fig. 5. Calculated barrier energies and attempt frequency prefactors for O-O, O-H, and H-O are listed in Table 3. The O-O path has a higher energy barrier than the O-H and H-

O paths, making O-O the path less favourable. The effective diffusion coefficient considering both paths can be calculated using diffusion equation (11).

$$D_{O \to O}^{Ti-0.01N} = \ell_{O \to O}^2 (\lambda_{O \to O} + \frac{3}{2} \lambda_{O \to H}) \tag{11}$$

where, $D_{O \to O}^{Ti-0.01N}$ is the diffusion coefficient of Ti-0.01N from O-O, $\ell_{O \to O}$ is the diffusion length of bulk Ti in O-O. $\lambda_{O \to O}$ and $\lambda_{O \to H}$ are the jump rates from the O-O and O-H, respectively. As the barrier for the O-H path is significantly lower than the O-O path, diffusion of N could be dominated by the O-H-O path.

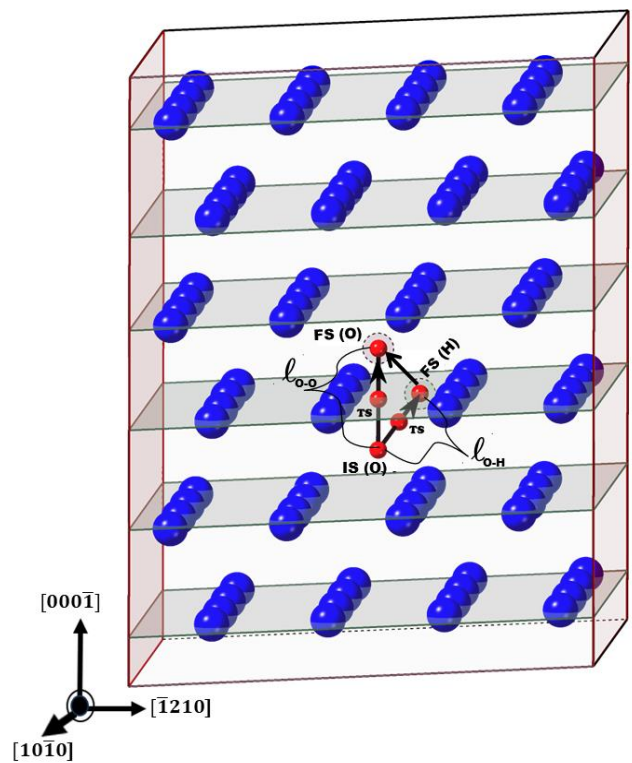

Fig. 5. Supercell representation of single N atom diffusion in HCP Ti along $[000\bar{1}]$ direction in two paths. one is O-O path, and the other O-H-O diffusion path.

### 3.2.4 N diffusion in Ti-0.20N

Diffusion of N along $[000\bar{1}]$ direction in Ti-0.20N was mediated by the N atom moving to the vacant octahedral site in the adjacent closed-pack plane. The diffusivity of the N atom can depend on the number and distribution of other N atoms at the neighbouring octahedral sites of the initial and final position of the diffusing N atom. We considered four different combinations of the initial and final position of the N atom, as shown in Fig. 6. The initial and final positions of N have different numbers and distributions from neighbouring N atoms. In all cases, we found that the hexahedral position was stable. So, we considered two transition networks: (1) direct O to O and (2) O to H to O. Fig. 6 shows four types of neighbouring environments for N hopping in adjacent planes through O-O, O-H and H-O pathways by considering initial and final positions as (i) all nearest neighbours were unoccupied (6-Vac), (ii) out of six, one nearest neighbour was occupied (5-Vac 1-Occ), (iii) out of six, two nearest neighbours were occupied (4-Vac 2-Occ), and (iv) out of six, three nearest neighbours were occupied (3-Vac 3-Occ). Here, Vac = Vacancy, Occ = Occupied. We took an average of the barrier energies and attempt frequency prefactors for four different transition choices, and barrier energies were listed in Table 2. The effective diffusion coefficient of N along the [000$\bar{1}$], in Ti-0.20N solid solution for O-O and O-H-O transition is given by equation (12).

$$D_{O \to O}^{Ti-0.20N} = \ell_{O \to O}^2 (\lambda_{O \to O} + \frac{3}{2} \lambda_{O \to H}) \qquad (12)$$

where, $D_{O \to O}^{Ti-0.20N}$ is the effective diffusion coefficient of the Ti-0.20N from O-O. $\lambda_{O \to O}$ and $\lambda_{O \to H}$ are the rate jumps from O-O and O- H, respectively. As the barrier for the O-H path was significantly lower than the O-O path, diffusion of N could be dominated by the O-H-O path.

Table 2. List of pathways of N diffusion in Ti-0.20N solid solution. The environment surrounding the initial and final position of the N atom. Barriers to various transition networks.

| Paths | Initial neighbouring environment | Final neighbouring environment | O-O (eV) | O-H (eV) | H-O (eV) |
|---|---|---|---|---|---|
| 1 | 4-Vac 2-Occ | 6-Vac | 3.30 | 2.92 | 0.45 |
| 2 | 3-Vac 3-Occ | 5-Vac 1-Occ | 3.43 | 1.81 | 0.72 |
| 3 | 5-Vac 1-Occ | 4-Vac 2-Occ | 3.32 | 2.26 | 0.55 |
| 4 | 6-Vac | 3-Vac 3-Occ | 3.32 | 2.25 | 0.56 |

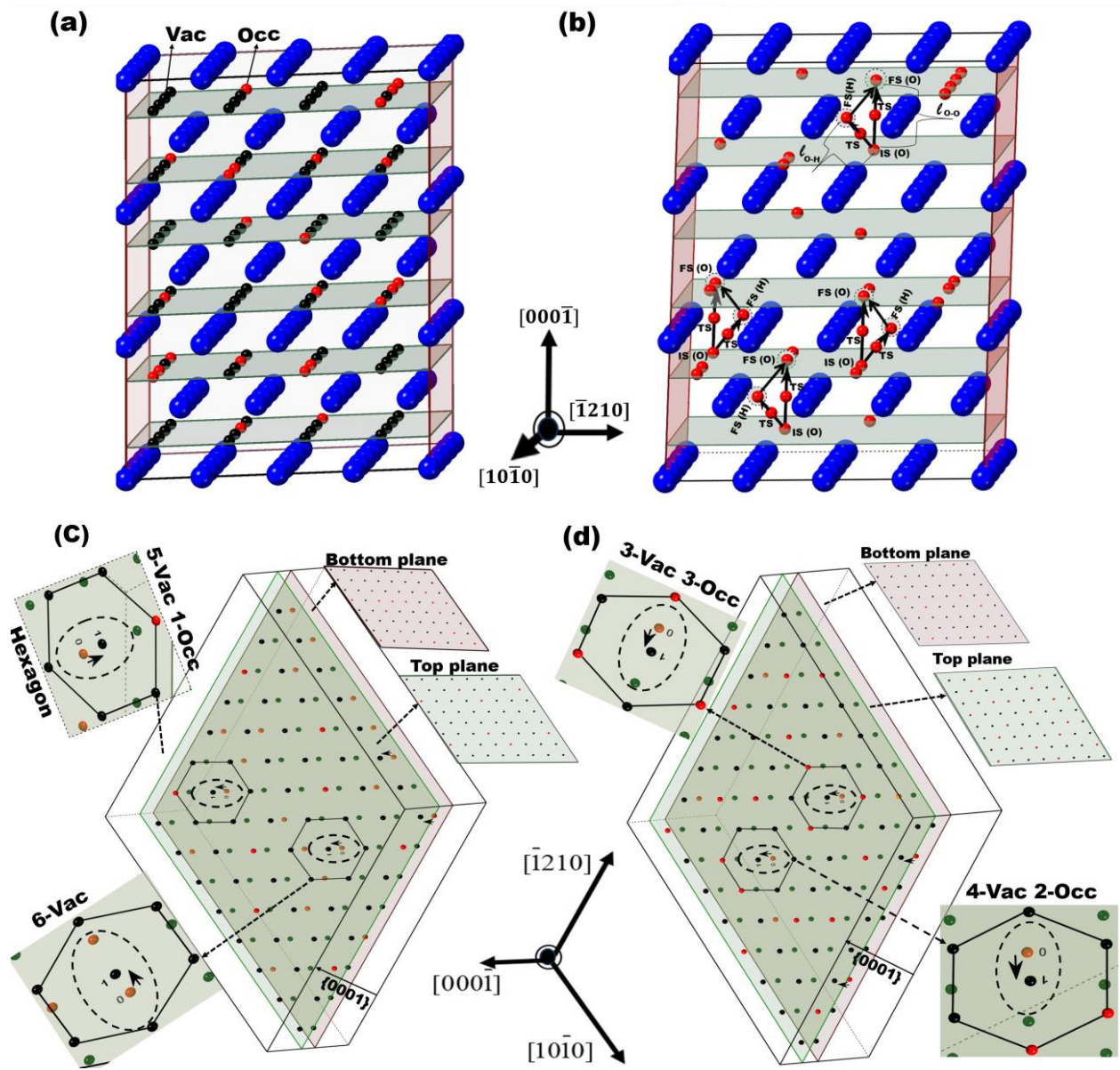

Fig. 6 Ti-0.20N supercell: (a) representing vacancies (Black spheres) and occupancy of N (Red spheres). (b) representing all pathways of N diffusions from O-O, O-H and H-O. (b&c) representation of several possible hexagonal arrangements of N atoms along $[000\bar{1}]$ direction, with N vacant positions in Ti-0.20N. 0 represents the initial position of N, and 1 represents the final position of N.

### 3.2.5 *Last N atom diffusion to form Ti₂N*

We assumed that Ti$_2$N forms layer by layer by filling half of the octahedral site between two Ti closed-packed planes. Diffusivity was calculated for the last N atom hopping from the octahedral site in the Ti layer above the near Ti$_2$N layer to the vacant octahedral site in the near

Ti₂N layer, as shown in Fig. 7. This process of diffusion completes the formation of Ti$_2$N, as shown in Fig. 7. We used a Ti/Ti$_2$N heterostructure supercell to calculate the diffusivity of N (Fig.7). The bottom of the supercell structure was Ti$_2$N, consisting of four Ti-layers and four N-layers, with the last N-layer having one N vacancy. The N diffusion path was assumed to be from the octahedral site in Ti to the vacant site in Ti$_2$N. We found that the hexahedral site (Fig. 7) was stable. So, we considered two transition networks: (1) direct O to O, and (2) O to H to O. We found that O to O path was unstable. Hence, we further considered only the O to H to O path. The energy barrier and attempt frequency prefactor for the O to H path are listed in Table 3. The effective diffusion coefficient of the last N atom into Ti$_2$N for the octahedral to hexahedral to octahedral transition is given by equation (13).

$$D_{O \to O}^{Ti/Ti_2N} = \ell_{O \to O}^2 (\frac{3}{2}\lambda_{O \to H}) \tag{13}$$

where, $D_{O \to O}^{Ti/Ti_2N}$ is the diffusion coefficients of Ti/Ti$_2$N from O-O. $\lambda_{O \to H}$ is the jump rate from O-H, $\ell_{O \to O}$ the diffusion length of the last N diffuses from O-H-O into Ti$_2$N. The corresponding diffusivity value is listed in Table 3.
.

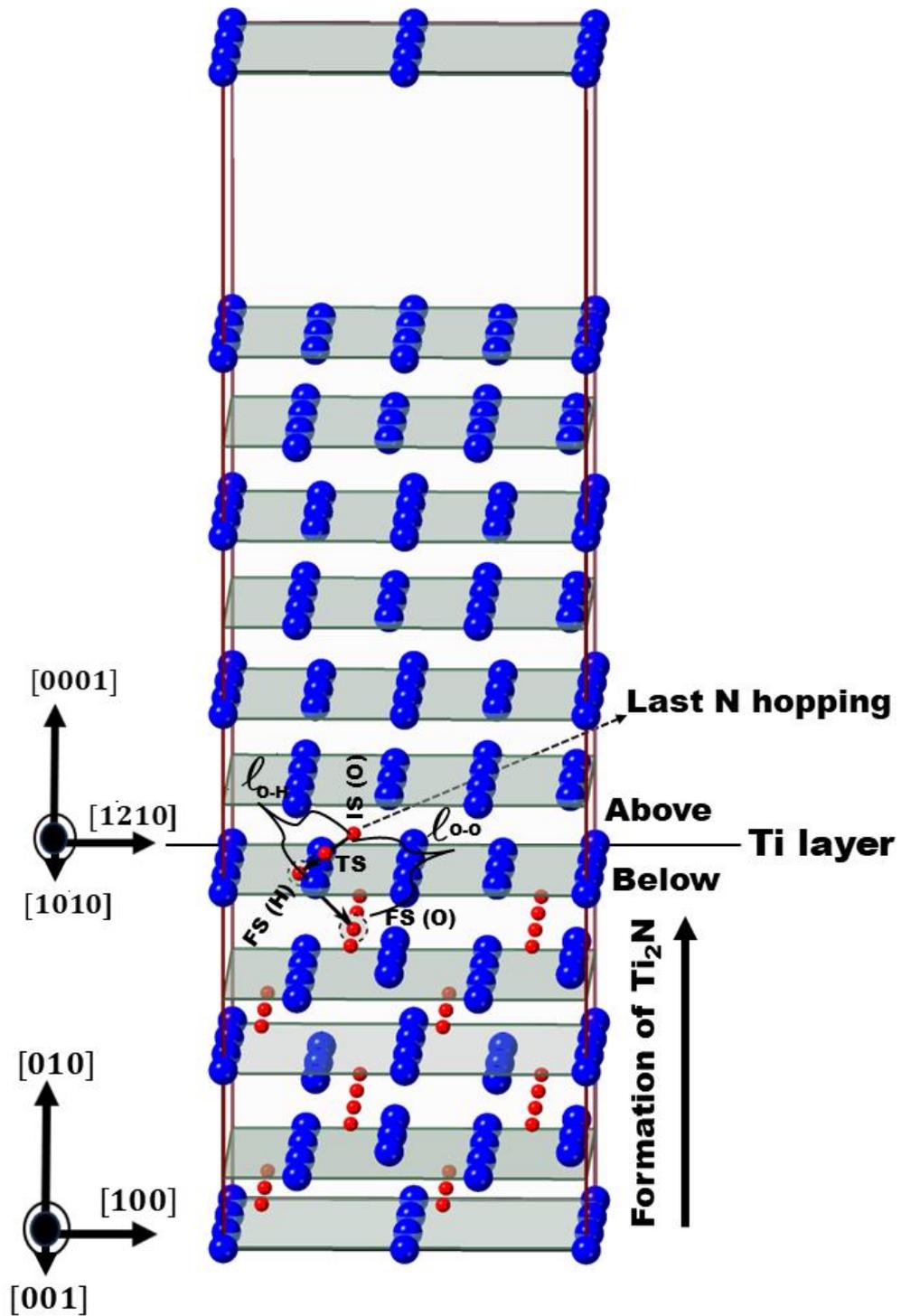

Fig. 7 The supercell representation of the last N atom intercalation to form Ti$_2$N in the Ti/Ti$_2$N system. N atom diffuses from Ti octahedral site to hexahedral site to octahedral site (O-H-O) in Ti$_2$N.

Table 3 lists the diffusivity of N at room temperature for all the above cases, i. e.: (1) from surface to sub-surface, (2) sub-surface to bulk, (3) Ti-0.01N, (4) Ti-0.20N and (5) at the last step of N intercalation to form Ti$_2$N. The rate of N diffusion from surface to sub-surface was faster than in the bulk. Diffusion coefficients of N in bulk remained the same as N concentration increased in Ti till the formation of $Ti_2N$. Except for the diffusion of N from surface to sub-surface, where the hexahedral site was not stable, all other paths favoured the diffusion of N along the octahedral to hexahedral to octahedral transition network. Thus, it was observed that the kinetics also supported the proposed process.

Table 3. List of the transition networks, perfectors, and energy barriers for N in various intermediate steps. The diffusivity of N from the surface to the bulk region to the formation of $Ti_2N$ is calculated at room temperature (RT).

| Transition Networks | | Barrier (eV) | Perfector (THz) | Diffusivity at RT ($m^2$/s) |
|---|---|---|---|---|
| Surface to sub-surface | | | | |
| FCC-O | FCC-O | 1.26 | 11.12 | 3.49 X $10^{-28}$ |
| Sub-surface to Bulk | | | | |
| O-O | O-O | 3.50 | 20.12 | 1.29 X $10^{-23}$ |
| O-H-O | O-H | 1.73 | 19.01 | |
| | H-O | 0.48 | 13.50 | |
| Ti-0.01N | | | | |
| O-O | O-O | 3.65 | 16.38 | 2.51 X $10^{-45}$ |
| O-H-O | O-H | 2.30 | 15.94 | |
| | H-O | 0.58 | 10.61 | |
| Ti-0.20 N | | | | |
| O-O (average) | O-O | 3.30 | 14.20 | 1.63 X $10^{-45}$ |
| O-H-O (average) | O-H | 2.31 | 13.94 | |
| | H-O | 0.42 | 10.20 | |
| N intercalation to form $Ti_2N$ | | | | |
| O-H-O | O-H | 2.33 | 13.60 | 7.07 X $10^{-46}$ |
| | H-O | 0.89 | 10.80 | |

## 4. Conclusions

The present study investigated the possibility of the formation of $Ti_2N$ by a nucleation-free growth process. Thermodynamic stability of different intermediate structures and the diffusion rates for the N atom at various stages were analyzed to understand the N intercalation from the Ti {0001} surface to the subsurface, leading to the formation of the $Ti_2N$ structure. The $N_2$ molecule adsorbed favorably onto the Ti surface and dissociated into N atoms, which preferred the stable FCC sites on the Ti surface. These N atoms were even more stable in the sub-surface. As N concentration in bulk Ti increased, Ti-N solid solution first formed, followed by $Ti_2N$ formation as the most stable structure. The diffusivity of N at various steps of nitridation up to the formation of $Ti_2N$ was also calculated. The diffusivity of N was relatively high when N diffused from the surface to the sub-surface but decreased as N moved into the bulk-like region. The diffusivity of N in bulk Ti was similar to the diffusivity of N in $Ti_2N$. It was well established that N diffusion in Ti is fast enough to form an N solid solution with Ti at high temperatures. Hence, we expected a similar rate of Ti nitriding to form $Ti_2N$. We concluded that Ti exposed to low concertation of N could form $Ti_2N$ by direct diffusion of N in Ti.

**Acknowledgments**

We acknowledge the use of the computing resources at High Performance Computing Environment (HPCE), IIT Madras. This work was supported by the Science and Engineering Research Board (SERB), New Delhi (Grant number: SRG/2019/000455) and the Ministry of Human Resource and Development (Gran number: SB20210844MMMHRD008277)

## Author Contributions

**J. Varalakshmi:** Conceptualization, methodology, formal analysis, writing-original draft, visualization, and theoretical computations were performed.
**Satyesh Kumar Yadav:** Conceptualization, resources, review & editing, supervision, funding acquisition.

## Data Availability

All data generated or analyzed during this study are included in this published article. Data is available upon reasonable request.

## Additional Information

**Competing financial interests**: The authors declare no competing financial interests.